\documentclass[12pt,preprint]{aastex}

\newcommand{\Swift}{{\it Swift}}
\newcommand{\Epsrc}{$E_{\rm peak}^{\rm src}$}
\newcommand{\Epobs}{$E_{\rm peak}^{\rm obs}$}
\newcommand{\Eiso}{$E_{\rm iso}$}
\newcommand{\Egamma}{$E_{\gamma}$}
\newcommand{\nuFnu}{$\nu F_{\nu}$}

\slugcomment{Accepted for publication in \apj}

\shorttitle{No Jet Break Feature in X-ray Band}
\shortauthors{Sato et al.}

\begin{document}


\title{\Swift~Discovery of Gamma-ray Bursts without Jet Break Feature in their X-ray Afterglows}

\author{
G.~Sato\altaffilmark{1,2},
R.~Yamazaki\altaffilmark{3}, 
K.~Ioka\altaffilmark{4},
T.~Sakamoto\altaffilmark{1,5},
T.~Takahashi\altaffilmark{2,6},
K.~Nakazawa\altaffilmark{2},
T.~Nakamura\altaffilmark{4},
K.~Toma\altaffilmark{4},
D.~Hullinger\altaffilmark{1,7,8},
M.~Tashiro\altaffilmark{9},
A.~M.~Parsons\altaffilmark{1},
H.~A.~Krimm\altaffilmark{1,10},
S.~D.~Barthelmy\altaffilmark{1},
N.~Gehrels\altaffilmark{1},
D.~N.~Burrows\altaffilmark{11},
P.~T.~O'Brien\altaffilmark{12},
J.~P.~Osborne\altaffilmark{12},
G.~Chincarini\altaffilmark{13,14},
and D.~Q.~Lamb\altaffilmark{15}
}

\altaffiltext{1}{NASA Goddard Space Flight Center, Greenbelt, MD 20771, USA}
\altaffiltext{2}{Institute of Space and Astronautical Science JAXA,
                 Kanagawa 229-8510, Japan}
\altaffiltext{3}{Department of Physics, Hiroshima University, 
                 Higashi-Hiroshima 739-8526, Japan}
\altaffiltext{4}{Department of Physics, Kyoto University, 
                 Kyoto 606-8502, Japan}
\altaffiltext{5}{Oak Ridge Associated Universities, P. O. Box 117, Oak Ridge, TN 37831, USA}
\altaffiltext{6}{Department of Physics, University of Tokyo, Bunkyo, 
                 Tokyo 113-0033, Japan}
\altaffiltext{7}{Department of Physics, Brigham Young University, Rexburg, ID 83440, USA}
\altaffiltext{8}{Department of Physics, University of Maryland, College Park, MD 20742, USA}
\altaffiltext{9}{Department of Physics, Saitama University, Saitama 338-8570, Japan}
\altaffiltext{10}{Universities Space Research Association, Columbia, MD 20744, USA}
\altaffiltext{11}{Department of Physics, Pennsylvania State University, University Park, PA 16802, USA}
\altaffiltext{12}{Department of Physics and Astronomy, University of Leicester, Leicester, LE 1 7RH, UK}
\altaffiltext{13}{Universit\`{a} degli studi di Milano-Bicocca, Dipartimento di Fisica, Italy}
\altaffiltext{14}{INAF-Osservatorio Astronomico di Brera, Via E. Bianchi 46, 23807 Merate(LC), Italy}
\altaffiltext{15}{Department of Astronomy and Astrophysics, University of Chicago, Chicago, IL 60637, USA}

\begin{abstract}
We analyze \Swift~gamma-ray bursts (GRBs) and X-ray afterglows for
three GRBs with spectroscopic redshift determinations --- GRB~050401,
XRF~050416a, and GRB~050525a.  We find that the relation between
spectral peak energy and isotropic energy of prompt emissions (the
Amati relation) is consistent with that for the bursts observed in
pre-\Swift~era.  However, we find that the X-ray afterglow 
lightcurves, which extend up to 10--70 days, show no sign of the jet
break that is expected in the standard framework of collimated
outflows.  We do so by showing that none of the X-ray afterglow
lightcurves in our sample satisfies the relation between the spectral
and temporal indices that is predicted for the phase after jet break. 
The jet break time can be predicted by inverting the tight empirical
relation between the peak energy of the spectrum and the
collimation-corrected energy of the prompt emission (the Ghirlanda
relation).  We find that there are no temporal breaks within the
predicted time intervals in X-ray band.  This requires either that the
Ghirlanda relation has a larger scatter than previously thought, that
the temporal break in X-rays is masked by some additional source of
X-ray emission, or that it does not happen because of some unknown 
reason. 
\end{abstract}


\section{Introduction}

The \Swift~satellite \citep{geh04} has enabled the acquisition of
early, dense, and detailed data on the X-ray afterglows of gamma-ray
bursts (GRBs) \citep[e.g.,][]{bur05a, tag05, obr06}.
Analysis of X-ray telescope \citep[XRT;][]{bur05b} data has revealed 
complex temporal behavior in the early phase of the afterglow
\citep{nou06, zha06, obr06}.
In addition to investigating the early phase of the X-ray afterglows, 
we can study the temporal and spectral properties of the X-ray
afterglows  at later times ($\gtrsim10^4$~s), which had been studied
mainly  using optical data before the \Swift~era.
It is widely believed that the GRBs arise from collimated outflows
(i.e., jets).  This picture is supported by the break from a shallower
to a steeper slope that is observed in many afterglow light curves at
around a day after the burst \citep{sari99}.   These breaks are
interpreted as being due to the geometrical effect caused by the
inverse of the bulk Lorentz factor of the jet becoming larger than the
physical opening angle of the jet, and  to a hydrodynamical transition
of the jet (i.e., a broadening of the jet), which is expected to occur
shortly afterward. The break is therefore expected to be independent of
wavelength (i.e., achromatic).
Importantly, in the standard synchrotron-shock model \citep{sari98},
the observed flux above the cooling frequency does not depend on the
density of ambient matter.  Consequently, the X-ray afterglow is 
expected to be less variable than the optical one.  Hence, observations
of X-ray afterglows are a useful tool for studying the jet break.

In this paper, we investigate the presence or absence of a jet break 
in the X-ray afterglows of recent \Swift~GRBs.  According to
\citet{fra01} and \citet{blo03}, given the observed jet break time, we
can calculate the jet opening angle and thereby the
collimation-corrected gamma-ray energy (\Egamma).  After correcting for
the jet collimation, \Egamma~shows a tight correlation with the peak
energy \Epsrc~of the \nuFnu~spectrum in the source-frame:  $E_{\rm
peak}^{\rm src} \propto E_{\gamma}^{0.7}$ \citep[the Ghirlanda
relation:][]{ghi04b}.
For \Swift~GRBs in which one can obtain both \Epsrc~and the
isotropic-equivalent gamma-ray energy  $E_{\rm iso} =
E_{\gamma}/(1-\cos\theta_{\rm j})$, where $\theta_{\rm j}$ is the
opening-half angle of the jet, the Ghirlanda relation can be inverted
to predict the value of $\theta_{\rm j}$, and hence the jet break time.
The X-ray afterglow can then be investigated to find out whether a jet
break is present at the expected epoch.  Hence, we can check the
validity of the Ghirlanda relation found for pre-\Swift~bursts using
mainly optical observations, and also the validity of the theory of the
jet break established in the pre-\Swift~era.

\section{Data Analysis}

\subsection{Data Selection}
In order to be able to do the analysis described above,
both prompt and afterglow data are necessary.
Among the 10 \Swift~long GRBs with measured redshifts detected before July 2005,
we find, for seven of them, either that the peak energy is hard to constrain
or that the XRT light curve was not observed for long enough.
We have thus selected the other three well-sampled bursts
(GRB~050401, XRF~050416a, and GRB~050525a) for our study.

The prompt emission of GRBs has a spectrum  that is  well described by
the Band function \citep{ban93}. We calculate the ``bolometric''
isotropic-equivalent gamma-ray energy, \Eiso, in the source frame by 
integrating the best-fit model for the time-averaged spectrum over the 
energy range 1--$10^4$~keV.  In order to do this, it is necessary to
know the overall shape of the spectrum, and therefore the three
parameters of the Band function.
In the cases of GRB~050525a~and XRF~050416a, we find that the peak
energy, \Epobs, of the gamma-ray spectrum falls within the energy range
of \Swift~Burst Alert Telescope \citep[BAT; 15--150~keV;][]{bart05}. 
The Band function gives a significantly better fit than does a single
power-law (PL) model or a power-law times exponential (PLE) model, and
adequately describes the BAT spectral data for these two bursts.  In
the case of GRB~050401, \Epobs~falls outside the energy range of BAT.
Since GRB~050401 was simultaneously observed \citep{gol05a} by 
Konus-Wind \citep[20~keV--14~MeV;][]{apt95}, we utilize the Konus-Wind 
spectral data to find \Epobs.

Here we describe the results of the spectral and temporal analyses that
we performed on the prompt emission and X-ray afterglow, of each
burst.  Table \ref{tab:sample} gives the redshift, $T_{90}$ duration,
peak photon energy flux, and photon energy fluence determined from our
analyses, while Table \ref{tab:parms} summarizes the best-fit spectral
parameters for the prompt emission and the X-ray afterglow for each
burst.
Throughout this paper, we use HEAsoft 6.0, which includes the
\Swift~software package (release 2005-08-08).  We also adopt a
cosmological model with $\Omega_{\rm m} = 0.3$, $\Omega_{\rm \Lambda} =
0.7$, and $H_0 = 70~{\rm km s}^{-1} {\rm Mpc}^{-1}$.
Errors quoted are at the 90\% confidence level unless otherwise stated.

\subsection{GRB~050401}

GRB~050401 was detected and localized by the \Swift~BAT at 14:20:15 UTC
on 2005 April 1 \citep{barb05}. \Swift~autonomously slewed to the GRB
position and the XRT found the X-ray afterglow emission at (R.A., Dec.)
=  ($16^{\rm h} 31^{\rm m} 28.85^{\rm s}, +02^{\circ}11'14.4''$)  with
the 90\% error radius of $3.3''$ \citep{mor06}.
The optical afterglow emission was also detected with several ground
observations \citep[e.g.,][]{ryk05}.  \citet{fyn05} detected several absorption
lines consistent with absorption systems at redshifts $z = 2.50$ and $z = 2.90$.
Following these authors, we adopt $z = 2.90$.

We first analyze the \Swift~BAT data for the prompt emission.  We
subtracted the background using the modulations of the coded aperture
mask (mask-weight technique).  The prompt emission has a $T_{90}$
duration of 34.3 s. The \Swift~BAT time-averaged spectral data in
15--150~keV is adequately fit by a single power-law model ($N(E)\propto
E^{-\Gamma}$) and gives a photon index of $\Gamma =
1.54_{-0.07}^{+0.07}$ with $\chi_{\nu}^{2} = 0.73$ (58 dof).
Neither the Band function nor the cut-off power-law model improves the
fit significantly.
We then analyzed the time-averaged spectral data from Konus-Wind, which
has a wider energy range. We used the spectral data from an
adjacent time domain to subtract the background from the spectral data
during the burst.  We then fit to the data a power-law (PL) model, a
power-law times exponential model (PLE) and a Band function: $N(E)
\propto E^{\alpha_{\rm B}}\exp(-\frac{E}{E_0})~{\rm for}~E<(\alpha_{\rm
B}-\beta_{\rm B})E_0; ~\propto E^{\beta_{\rm B}}~{\rm
for}~E\geq(\alpha_{\rm B}-\beta_{\rm B})E_0$, where $\nu F(\nu)$ peaks
at $E_{\rm peak}^{\rm obs}=(\alpha_{\rm B}+2)E_0$. We find
$\chi_{\nu}^{2} = 2.38$ (58 dof), $\chi_{\nu}^{2} = 1.12$ (57 dof), and
$\chi_{\nu}^{2} = 1.00$ (56 dof), respectively.
Thus the spectral data strongly requests the Band function over the PL and PLE models.
The best fit values and uncertainties for the Band function
parameters obtained in this way are
\Epobs~$=115_{-16}^{+19}$~keV,
$\alpha_{\rm B}=-0.87_{-0.27}^{+0.36}$, and
$\beta_{\rm B}=-2.47_{-0.36}^{+0.21}$, respectively.
Using the redshift $z = 2.90$, the peak energy at the rest frame of the
GRB is determined as \Epsrc~$= 447_{-64}^{+75}$~keV, and the isotropic
energy as \Eiso~$= 3.43_{-0.34}^{+0.37} \times 10^{53}$ erg over 1 keV
to 10 MeV.

We next analyze the \Swift~XRT data for the event. The XRT acquired
data mainly in Windowed Timing (WT) mode in the first $\sim10000$ s
from the BAT trigger, and then switched to Photon Counting (PC) mode
according to the source count rate. We used XSELECT to extract source
and background counts from the cleaned event list (0.5--10.0 keV),
using the standard grade selections of 0--12 for PC mode data, and of
0--2 for WT mode data. We calculate the source light curve and spectrum
from a region with a  length of $80''$ in uncompressed direction for WT
mode, and a circular region with a radius of $47''$ for PC mode. We
extract the background light curves and spectra from outer regions,
excluding other X-ray sources that are visible in the XRT image. We
converted the count rate to the unabsorbed flux in the  2--10 keV
energy band using the best fit spectral model.

\begin{figure}
\plotone{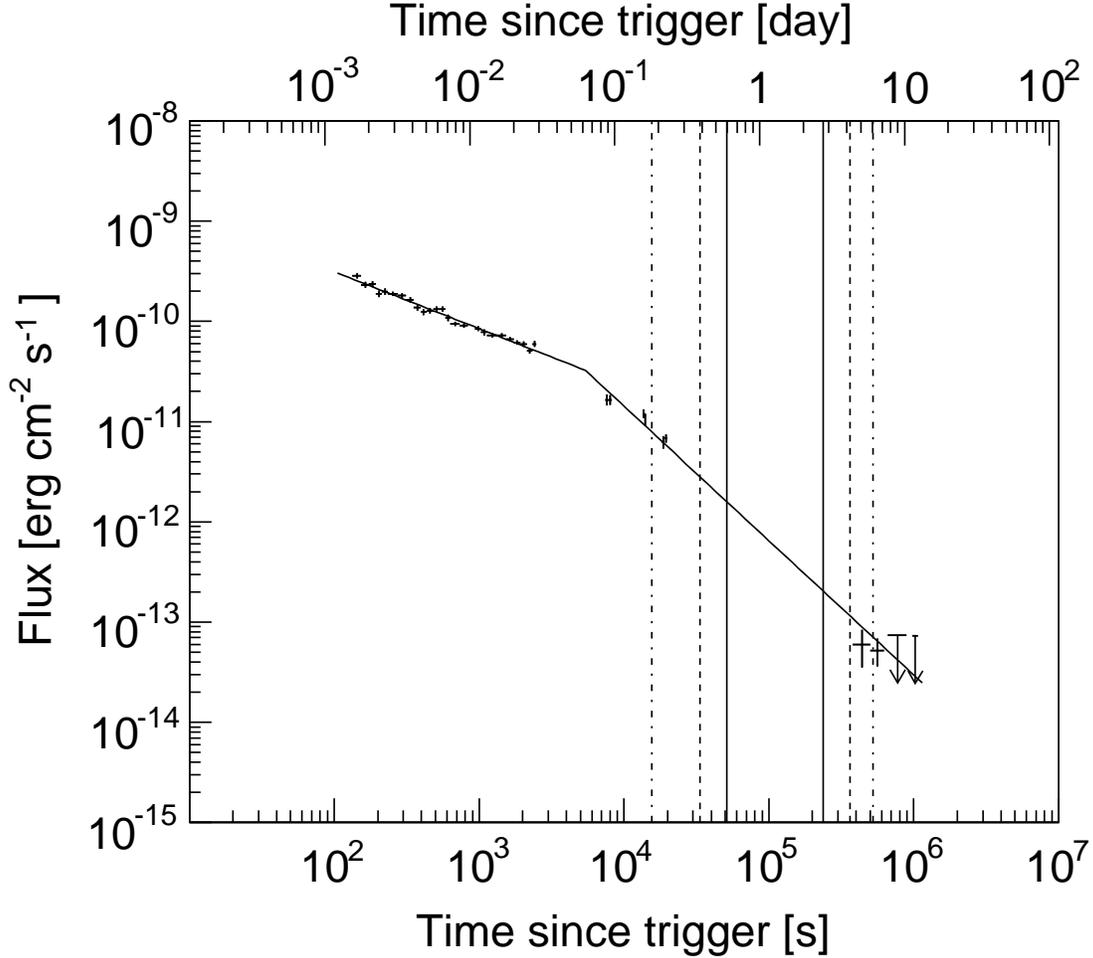}
\caption{X-ray afterglow light curve of GRB~050401 in the 2--10 keV
energy band.  In order to satisfy the Ghirlanda relation between \Epsrc
and \Egamma, the X-ray light curve should exhibit a jet break within
the time interval indicated by  the vertical lines.  The dash-dotted,
dashed, and solid  lines show, respectively, the allowed time intervals, without
assuming a particular value of $n\eta_\gamma$ and taking into account the
errors in \Eiso~and \Epsrc~in Eq.~\ref{eq:e3}; assuming a particular value of $n\eta_\gamma$ and
taking into account the errors in \Eiso~and \Epsrc; and assuming a
particular value of $n\eta_\gamma$ without taking into account the errors in
\Eiso~and \Epsrc.
Here, $n$ is the number density of the ambient (uniform) medium,
and $\eta_\gamma$ is the efficiency of the shock in
converting the energy in the ejecta into $\gamma$-rays.
See \S\ref{sec:dis1} for more explanations.
\label{fig:xlc1}}
\end{figure}

Fig.~\ref{fig:xlc1} shows the background-subtracted 2--10 keV light curve.
The X-ray afterglow of GRB~050401 faded slowly with a very shallow 
temporal index in the time interval from T+134 s to T+2484 s.  Here, T
represents the trigger time of the BAT. Extrapolating the initial slope
to late times, the WT data in T+7414 s -- T+8274 s and the PC mode
data in T+13486 s -- T+14066 s clearly have lower fluxes than expected.
The XRT also detected the fading afterglow at a later time between
T+4.4 days and T+ 7.2 days. Including these data points, the best fit
to the overall light curve is  given by a broken power-law model:
$F(t)\propto t^{\alpha_1}~{\rm for}~t<t_{\rm b};~\propto
t^{\alpha_2}~{\rm for}~t\geq t_{\rm b}$. The best-fit model gives
$\chi_{\nu}^{2} = 1.59$ (29 dof), with best-fit parameters $\alpha_{1}
= -0.57 \pm 0.02$, $\alpha_{2} = -1.34 \pm 0.05$ and $t_{\rm b} = 5390
\pm 450$ s.  This result is consistent with that of \citet{dep06}.
We have also analyzed the spectral data before and after the temporal
break. We find that both spectra are well-fit by a power-law model. The
best-fit model requests more absorption than the value $N_{\rm H} = 
4.9 \times 10^{20}$  cm$^{-2}$ expected for the galaxy alone.  We
have therefore added absorption at the redshift of the GRB
($z= 2.90$) to the model. The best-fit parameters are $\Gamma =
2.03_{-0.05}^{+0.05}$ and $N_{\rm H}^{z} = 3.7_{-0.5}^{+0.5} \times
10^{22}$ cm$^{-2}$ ($\chi_{\nu}^{2} = 1.15$ (241 dof)) prior to the
break, and $\Gamma = 1.98_{-0.24}^{+0.26}$ and $N_{\rm H}^{z} =
3.0_{-2.5}^{+3.1}\times10^{22}$ cm$^{-2}$ ($\chi_{\nu}^{2} = 0.65$ (22
dof)) after the break.  There is therefore no significant evidence for
evolution of the spectral shape from before the break to after it,
taking into account the uncertainties in the spectral parameters.

\subsection{XRF~050416a}

The X-ray flash, XRF~050416a, was detected and localized by the
\Swift~BAT at 11:04:44.5 UTC on 2005 April 16 \citep{sak05}.
\Swift~autonomously slewed to the GRB position and \Swift~XRT found the
X-ray afterglow emission at (R.A., Dec.) =  ($12^{\rm h} 33^{\rm m}
54.63^{\rm s}, +21^{\circ}03'27.3''$)  with the 90\% error radius of
$3.3''$ \citep{mor06}. The detailed analysis of the BAT, XRT and UVOT
data are reported in several papers (BAT; \citet{sak06a}, XRT;
\citet{man06}, UVOT; \citet{hol06}).  The spectrum of the host galaxy
of XRF 050416a was obtained using the 10 m Keck I telescope; the host
galaxy is faint and blue with a high star formation rate and
its redshift is $z = 0.6535 \pm 0.0002$ \citep{cen05}.

The prompt emission had a duration of $T_{90} = 2.4$~s. XRF~050416a is
the softest burst observed by \Swift~BAT as of July 2005. \citet{sak06a} showed
that the time-averaged spectrum is much steeper than the photon index
of $\Gamma=2$, indicating the spectral peak lies at the lower end of or
below the BAT energy range.
Following these authors, we adopt the Band function model with a fixed
$\alpha_{\rm B}=-1$, which is the typical value for BATSE GRBs \citep{kan06}.
The fit gives \Epobs~$=18.0_{-2.9}^{+3.9}$~keV.
In order to take into account the uncertainty in the low energy photon
index, which may affect the total isotropic energy, \Eiso, we have
performed spectral fits to the \Swift~BAT spectral data, varying
$\alpha_{\rm B}$ from $-1.5$ to $-0.67$.
These limits correspond to the indices predicted for a spectrum in the
fast cooling phase; i.e., with $\nu_{\rm c}<\nu<\nu_{\rm m}$ and
$\nu<\nu_{\rm c}$, respectively \citep{sari98}.  Here, $\nu_{\rm c}$ is
the synchrotron cooling frequency and $\nu_{\rm m}$ is the synchrotron
frequency of electrons with the minimum energy.
We then find the best-fit values of the Band function parameters and
their uncertainties to be \Epobs~$= 17.3_{-8.0}^{+3.0}$~keV and 
$\beta_{\rm B} < -3.35$ with $\chi_{\nu}^{2} = 0.80$ (56 dof).   Using
the observed redshift of $z = 0.6535$, we find \Epsrc~$=
28.5_{-13.2}^{+5.0}$~keV, and \Eiso~$= 8.3_{-1.3}^{+6.5} \times
10^{50}$ erg.

The XRT data were acquired in PC mode throughout the observation.  We
extracted the light curves and spectra from the data in a circular
region with a radius of $47''$. The data obtained in PC mode sometimes
suffered from pile-up, especially when the count rate was higher than
0.5 counts/s \citep{nou06}.  For this burst, the XRT count rate
exceeded this limit for the first $\sim500$ s of the observation.  From
image analysis, we find the central region with radius of $6''$
deviates from the XRT point spread function.  We therefore excluded the
events in this region when we derived the source light curve and
spectrum for the time interval T+94 -- T+596 s; we used the full region
of the circle with radius of $47''$ in the later period.  The effective
area was corrected using the calibration data and the {\it FTOOL}
xrtmkarf.

\begin{figure}
\plotone{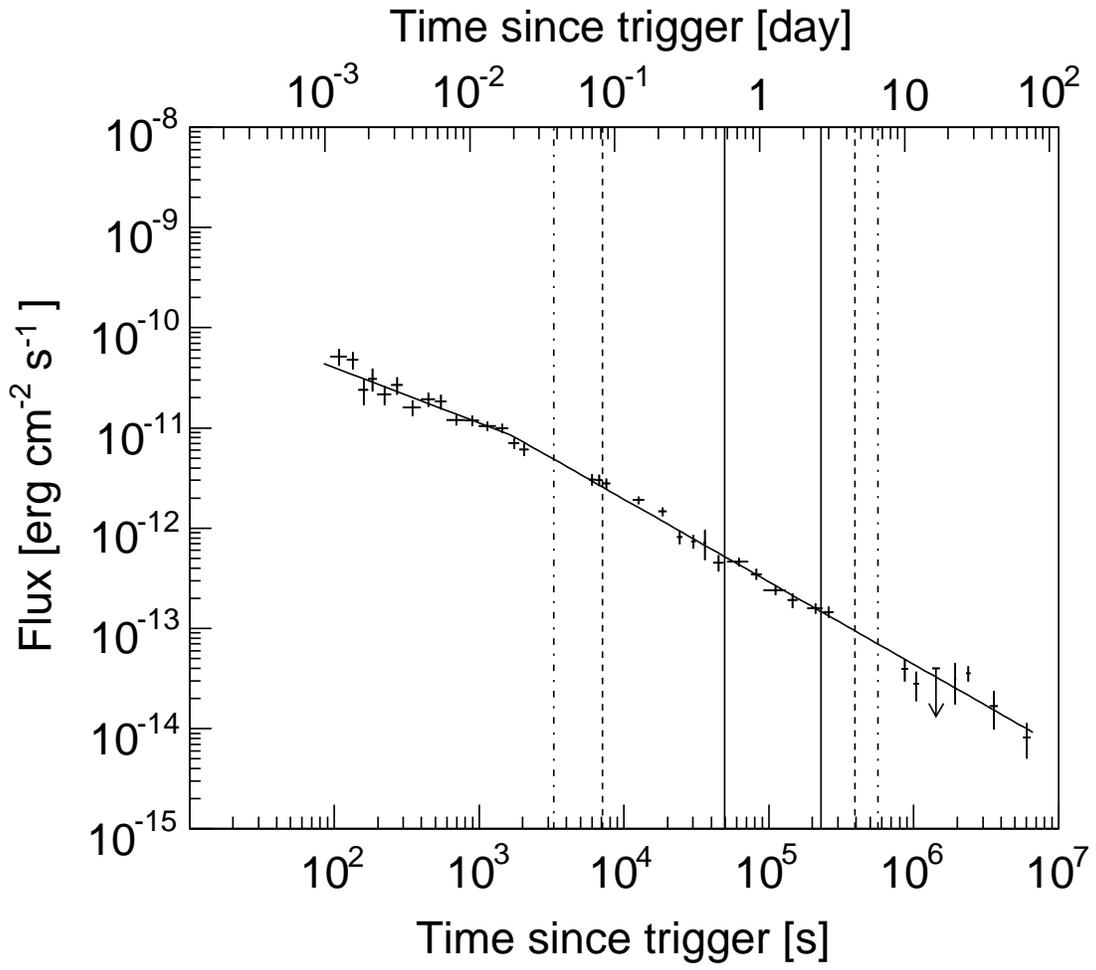}
\caption{The same as Fig.~\ref{fig:xlc1} but for XRF 050416a.
\label{fig:xlc2}}
\end{figure}

Fig.~\ref{fig:xlc2} shows the background-subtracted 2--10~keV light
curve  The light curve is well fit by a broken power-law with
$\chi_{\nu}^{2} =  1.12$ (32 dof). There is an indication of a break in
the light curve at $t_{\rm b} = 1670 \pm 600$~s, which is also reported
by \citet{nou06}.  The decay index is $\alpha_{1} = -0.55\pm0.06$ before the break
and $\alpha_{2} = -0.82\pm0.03$ after it. 
Strikingly, the light curve shows a shallow decay extending to
$\sim74.5$ days after the trigger.  The spectral data before and after
the break are both well fit by a power-law model with galactic
absorption ($N_{\rm H} = 3.4 \times 10^{20}$~cm$^{-2}$) and an
additional absorption component at the redshift of the GRB (z=
0.6535).  The best-fit parameters are $\Gamma = 2.20_{-0.24}^{+0.27}$
and $N_{\rm H}^{z} = 7.3_{-3.2}^{+3.7}\times10^{21}$~cm$^{-2}$
($\chi_{\nu}^{2} = 1.38$ (20 dof)) before the break, and $\Gamma =
2.04_{-0.15}^{+0.16}$ and  $N_{\rm H}^{z} = 5.5_{-1.9}^{+2.2} \times
10^{21}$~cm$^{-2}$ ($\chi_{\nu}^{2} = 0.93$ (64 dof)) after the break.
Thus, there is no significant evidence for spectral evolution, after
taking into account the uncertainties in the spectral parameters.

\subsection{GRB~050525a}

GRB~050525a was a very bright GRB that was detected and localized by
the \Swift~BAT at 00:02:52.8 UTC on 2005 May 25 \citep{ban05}.
\Swift~autonomously slewed to the GRB position, and \Swift~XRT and UVOT
started their observations about 100 s after the trigger,
and both found a fading source.
The optical coordinates are (R.A., Dec.) = 
($18^{\rm h} 32^{\rm m} 32.62^{\rm s}, +26^{\circ}20'21.6''$)
with an estimated uncertainty of $0.2''$. \citep{blu06}.
\citet{fol05} used GMOS on the Gemini-North telescope to obtain an
optical spectrum of the burst and reported that the redshift of the
host galaxy is  $z =0.606$ based on [O III] 5007 and H beta emission
and Ca H\&K and  Ca I 4228 absorption.

The prompt emission had a duration of $T_{90} = 8.9$ s. We fit the
\Swift-BAT time-averaged spectral data using a power-law (PL) model, a
power-law times exponential model (PLE) and a Band function.  We find
$\chi_{\nu}^{2} = 3.30$ (58 dof), $\chi_{\nu}^{2} = 0.26$ (57 dof), and
$\chi_{\nu}^{2} = 0.27$ (56 dof).
Thus both the PLE model and the Band function are acceptable.
We here employ the Band function
to constrain the upper limit on the higher energy index ($\beta_{\rm B}$)
and adequately include the uncertainty into the calculation of \Eiso.
The Band
function fit to the \Swift~BAT  time-averaged spectrum in 15--150 keV
gives best-fit parameters \Epobs~$= 78.2_{-1.6}^{+4.7}$~keV,
$\alpha_{\rm B} = -0.97_{-0.10}^{+0.11}$ and $\beta_{\rm B} < -2.55$.
Using the observed redshift of
$z = 0.606$, we find \Epsrc~$= 125.6_{-2.6}^{+7.6}$~keV, and \Eiso~$=
2.23_{-0.11}^{+0.03} \times 10^{52}$ erg.

\begin{figure}
\plotone{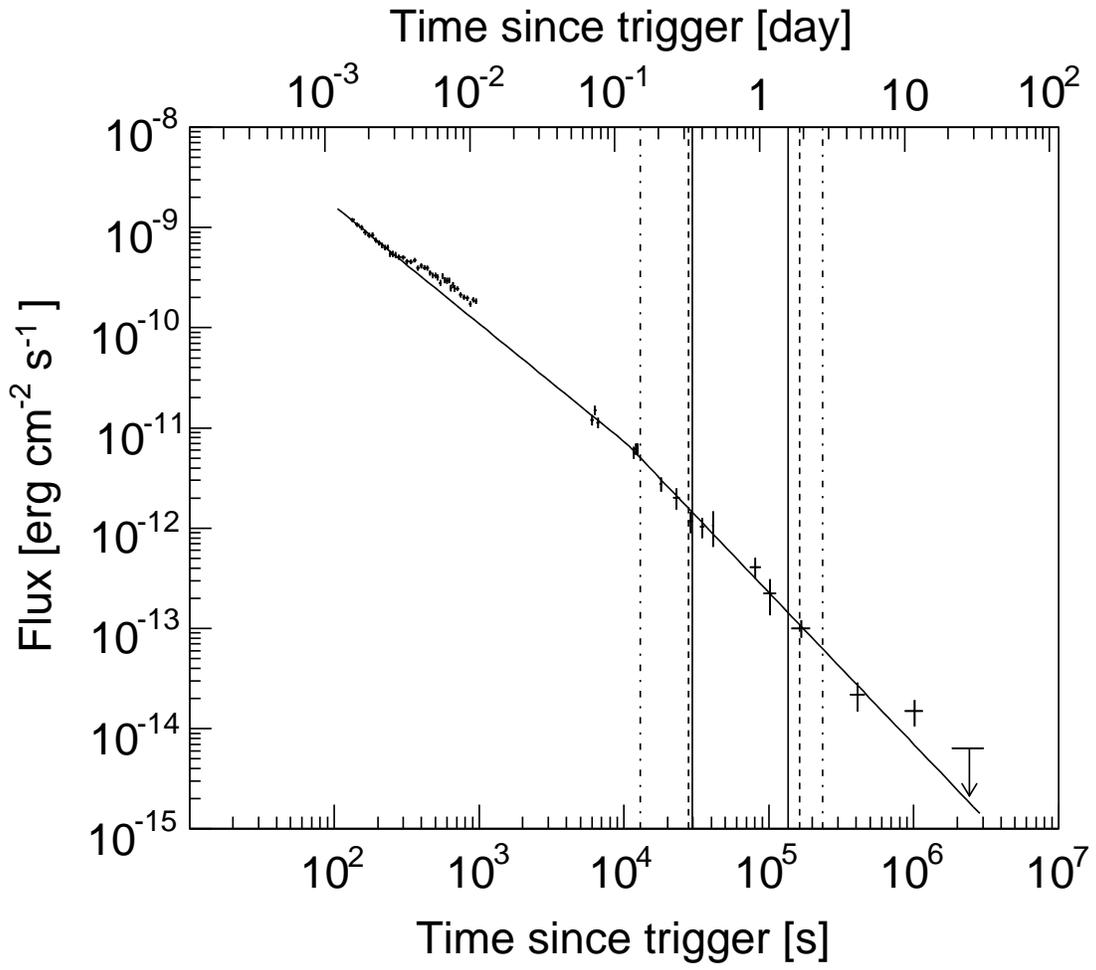}
\caption{The same as Fig.~\ref{fig:xlc1} but for GRB 050525a.
\label{fig:xlc3}}
\end{figure}

The X-ray afterglow was very bright just after the trigger.  Hence, the
observation was made first in Photo-Diode (PD) mode and was switched to
PC mode after T+5859 s.
No WT data were taken because of engineering tests that were being
performed at the time of the burst.
Since we cannot eliminate photons from the calibration source in PD
mode, we use PD data in the 0.5-4.5 keV band, while we use PC data in
0.5--10.0 keV.  Fig.~\ref{fig:xlc3} shows the background subtracted
2--10 keV light curve. We extrapolated the spectrum in PD mode to the
wider band when making the light curve. The last 2 points (T+10.0 days
-- T+35.0 days) are not considered in \citet{blu06} whose result for
$<$~T+5.4~days is consistent with ours.   In the time interval from
T+280~s to T+1048~s, we can see an excess above the fitted line.
Following \citet{blu06}, we identify the excess with a weak flare. If
we fit the data excluding the flare regime and the last upper limit, a
broken power-law model gives $\chi_{\nu}^{2} = 0.54$ (27 dof), with
best fit parameters $\alpha_{1} = -1.18 \pm 0.02$, $\alpha_{2} = -1.51
\pm 0.06$ and $t_{\rm b} = 10600 \pm 3300$ s.  The spectral data before
and after the break are extracted from the  entire PD data and the PC
data after break, respectively. Both spectra are well fit by a
power-law model with a galactic absorption ($N_{\rm H} = 9.1 \times
10^{20}$~cm$^{-2}$) and an additional absorption at the redshift of the
GRB (z= 0.606). The best-fit parameters are $\Gamma =
1.92_{-0.05}^{+0.05}$ and $N_{\rm H}^{z} =
2.6_{-0.4}^{+0.4}\times10^{21}$~cm$^{-2}$ ($\chi_{\nu}^{2} = 1.09$ (271
dof)) before the break, and $\Gamma = 2.11_{-0.39}^{+0.28}$ and $N_{\rm
H}^{z} = 1.5_{-1.5}^{+3.6}\times10^{21}$~cm$^{-2}$ ($\chi_{\nu}^{2} =
0.79$ (22 dof)) after the break.
Thus, there is no significant evidence for spectral evolution, after 
taking into account the uncertainties in the spectral parameters.

\section{Results and Discussion \label{sec:dis}}

\subsection{Investigation of Jet Break Features \label{sec:dis1}}

\begin{figure}
\epsscale{0.80}
\plotone{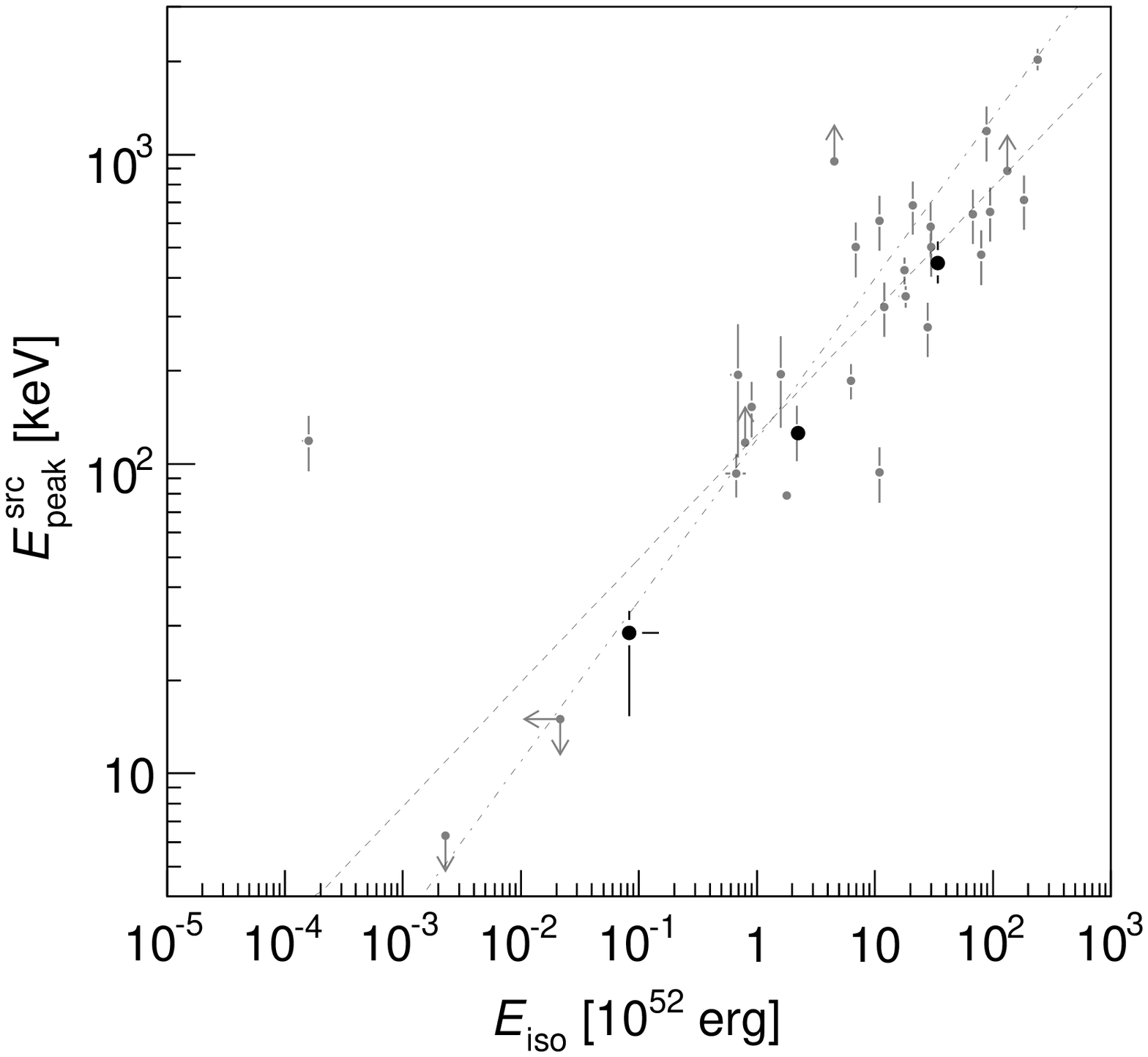}
\caption{Locations of GRBs in the (\Eiso,\Epsrc)-plane, where \Eiso~is
the isotropic-equivalent energy and \Epsrc~is the peak energy of the
burst spectrum in the rest frame of the burst.  The three filled black
circles correspond (from lower left to upper right) to XRF 050416a, GRB
050525a, and GRB 050401.  The burst locations previously  reported by
\citet{ghi04b} are shown as filled gray circles.  The dashed and
dot-dashed lines are the correlations between \Eiso~and \Epsrc~reported
by \citet{ama02} and \citet{ghi04b}, respectively.  The locations of
XRF~050416a, GRB~050525a, and GRB~050401, derived from Swift~and
Konus-Wind observations, lie within the scatter of the previous
\Eiso--\Epsrc~relation.
\label{fig:ama}}
\end{figure}

Spectral parameters of the prompt emission are well constrained by the 
\Swift~BAT and Konus data plus the optically determined redshifts.
Fig.~\ref{fig:ama} shows the locations of GRBs in the
(\Eiso,\Epsrc)-plane, where \Eiso~is the isotropic-equivalent energy
and \Epsrc~is the peak energy of the burst spectrum in the rest frame
of the burst. The burst locations previously reported by \citet{ghi04b}
are shown as filled gray circles.  The dashed and dot-dashed lines are
the correlations between \Eiso~and \Epsrc~reported by \citet{ama02} and
\citet{ghi04b}, respectively.  The locations of XRF~050416a,
GRB~050525a, and GRB~050401, derived from Swift~and Konus-Wind
observations, lie within the scatter of the Amati relation
\citep[$E_{\rm peak}^{\rm src} \propto E_{\rm iso}^{0.5}$:][]{ama02}.
Although it has been suggested that the Amati relation may have a large
intrinsic scatter [\citet{nak05,bp05}, but see \citet{ghi05,bos05}],
the locations of the three bursts discussed in this paper lie close
to the best-fit relations derived by \citet{ama02} and \citet{ghi04b}.

The X-ray follow-up observations for the three bursts start at
$\sim$T+100 s and end at T+12.4 -- T+74.5~days.  The X-ray afterglow
light curves do not exhibit a steep decline at the beginning of the 
observations, although \citet{man06} and \cite{obr06} show that a
fairly steep early decline can be seen for XRF~050416A by combining BAT
data with XRT image mode and low rate mode data. The light curves show
breaks at an early epoch, $t_{\rm b} \sim 10^3$--$10^4$~s which is similar
to the behavior seen in the X-ray afterglow of other bursts
\citep{nou06, obr06}. The decay slopes after the breaks are shallower
than the $\propto t^{-2}$ behavior expected after the jet-break
\citep{sari99,dai01}.

\begin{figure}
\epsscale{0.80}
\plotone{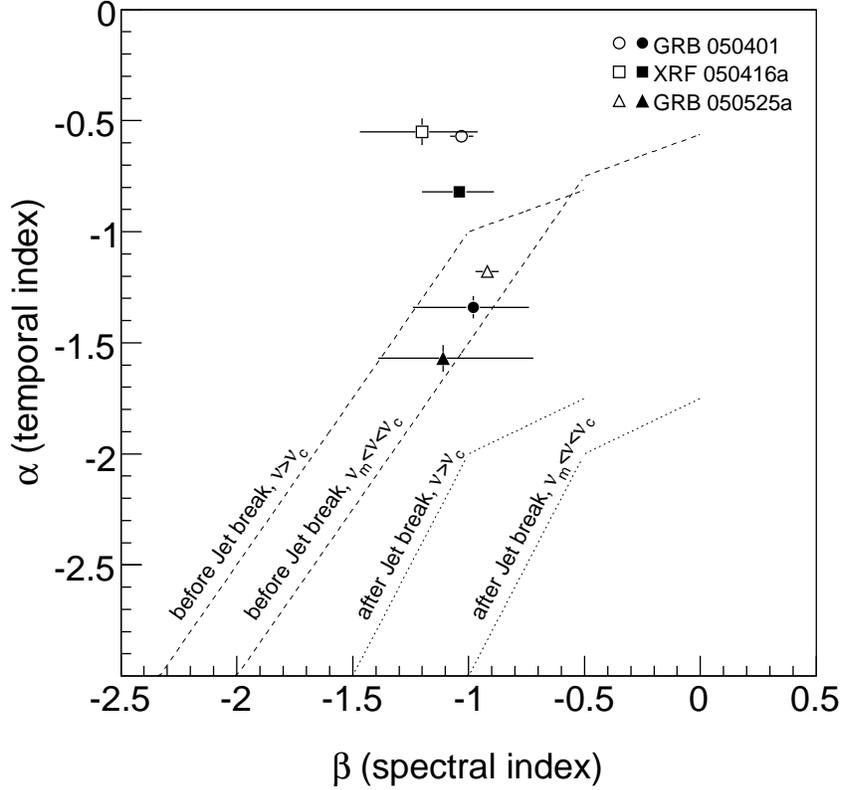}
\caption{Expected relation between the temporal index $\alpha$ and the
spectral index $\beta (= -\Gamma + 1)$, assuming a uniform density
(corresponding to an ISM environment) and that the external shock has
reached the slow-cooling phase.  The open symbols show the locations of
GRB 050401, XRF 050416a, and GRB 050525a, prior to the X-ray break at
$t_{\rm b}$,  while the closed symbols shows the locations of the three
bursts after the break.  None of the three bursts satisfy the
post-break relations expected in the standard afterglow model.
\label{fig:alpha_beta}}
\end{figure}

We first consider the behavior of the X-ray afterglows of the three
bursts  within the framework of the standard afterglow model in the
pre-\Swift~era.
The jet decelerates rapidly after the sideways expansion becomes 
significant, and the external shock enters the slow-cooling phase
\citep{sari98}. For example, if the the cooling frequency lies below
the X-ray band  (i.e., $\nu > \nu_{\rm c}$), the temporal decay index
($\alpha$) and the energy spectral index ($\beta=-\Gamma+1$) after the
jet break are given by $\alpha=-p$ and $\beta=-p/2$, respectively.
Here, $p > 2$ is the power-law index of the electron energy
distribution \citep{sari99}. Eliminating $p$, gives a relation between
$\alpha$ and $\beta$ (the so called $\alpha$--$\beta$ relation):
$\alpha=2\beta$.  If $1<p< 2$, the $\alpha$--$\beta$ relation takes a
different form: $\beta=2\alpha+3$ \citep{dai01}.
Similar formulae exist when the cooling frequency lies above the X-ray
band (i.e., $\nu_{\rm m}<\nu<\nu_{\rm c}$) but this is likely to be true
only at early times.
The observed results for the three bursts are shown in
Fig.~\ref{fig:alpha_beta} together with the theoretical predictions of
the $\alpha-\beta$ relations before and after the jet break, shown as
dashed and dotted lines respectively.
It is clear that none of the observed data points is consistent  with
the theoretical prediction of the standard afterglow models after the
jet break.

\begin{figure}
\epsscale{0.80}
\plotone{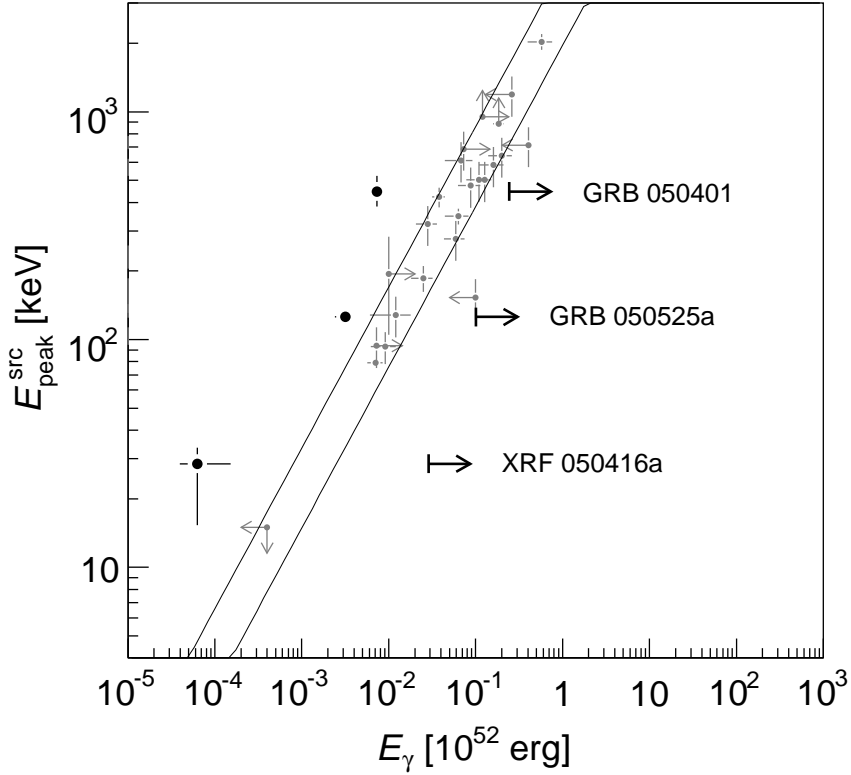}
\caption{Locations of GRBs in the ($E_\gamma$,\Epsrc)-plane, where 
$E_\gamma$ is the collimation-corrected energy and \Epsrc~is the peak 
energy of the burst spectrum in the rest frame of the burst.  The 
locations of the bursts in the samples of \citet{ghi04b}
are plotted as filled gray circles.  All of the bursts with 
well-constrained values of $E_\gamma$ and \Epsrc~in the \citet{ghi04b} 
sample lie inside the two solid diagonal lines corresponding to the Ghirlanda
relation (Eq. \ref{eq:e1}) calculated for $A = 1950$ keV and $A = 4380$
keV.  In the cases of GRB~050401, XRF~050416a, and GRB~050525a, no jet break
was observed in the X-ray afterglow light curve at an epoch which would 
allow $E_\gamma$ and \Epsrc~to lie in the band between the two solid lines.  
The filled black circles show the locations of the three
bursts, assuming that the X-ray break observed at an early time $t_{\rm b}$ 
is, in fact, the jet break.  The lower limits on $E_\gamma$ for the
three bursts assume that the jet break occurs after the last Swift
XRT observation of the X-ray afterglow.  In either case, \Egamma~lies 
outside of the band defined by the two diagonal solid lines for all the three \Swift~bursts.
\label{fig:ghi}}
\end{figure}

For each of the three events, we find no clear evidence of a jet break
in the X-ray light curve earlier than $10^6$ sec after the trigger.
We therefore invert the Ghirlanda relation to predict the jet break time
for each of the three bursts that makes them satisfy the
\Epsrc--\Egamma~relation.  The Ghirlanda relation is \citep{ghi04b} 
\begin{equation}
E_{\rm peak}^{\rm src} = AE_{\gamma, 52}{}^{0.706}~~, \label{eq:e1}
\end{equation}
where $E_{\gamma}=(1-\cos\theta_{\rm j})E_{\rm iso}$ is the 
collimation-corrected energy and $\theta_{\rm j}$ is the opening
half-angle of the jet. Here we define $E_{\gamma, 52} =
E_\gamma/10^{52}$ erg.
The relation is based on the jet breaks observed mainly in the optical
band.  However, the jet break should appear in the X-ray band at the same
time that it appears in the optical band because the break is
geometrical and hydrodynamical in origin.
Fig.~\ref{fig:ghi} shows the correlation between \Egamma~and
\Epsrc~for the GRB samples in \citet{ghi04b}.  The left and right
diagonal lines in the figure are for $A=4380$~keV and $A=1950$~keV, 
respectively; the band between the two lines includes all of the
central values of the locations of GRBs with well-constrained
properties in the \citet{ghi04b} samples.
The expression for the jet break time is  \citep{sari99}
\begin{eqnarray}
t_{\rm jet} = 130~\theta_{\rm j}^{8/3} (1+z)
\left(\frac{n\eta_{\gamma}}{E_{{\rm iso},52}}\right)^{-1/3}~~~~~{\rm days}, \label{eq:e2}
\end{eqnarray}
where $E_{\rm iso, 52}$, $n$, $\eta_\gamma$, and $z$ are respectively the
isotropic-equivalent energy in units of $10^{52}$ erg, the number
density of the ambient (uniform) medium, the efficiency of the shock in
converting the energy in the ejecta into $\gamma$-rays, and the source
redshift. Using Eqs.~(\ref{eq:e1}) and (\ref{eq:e2}), we
obtain
\begin{eqnarray}
t_{\rm jet} = 389~(1+z)\left(\frac{n}{3~{\rm cm}^{-3}}\right)^{-1/3}\left(\frac{\eta_\gamma}{0.2}\right)^{-1/3}
E_{{\rm iso},52}^{-1}\left(\frac{E_{\rm peak}^{\rm src}}{A}\right)^{1.89}~~~~~{\rm days}.
\label{eq:e3}
\end{eqnarray}
Using this equation, we can calculate $t_{\rm jet}$ from the  $E_{\rm
peak}^{\rm src}$~and \Eiso~values we have derived from the
observations.  The efficiency $\eta_\gamma$ of the shock, and
especially, the number density $n$ of the ambient medium, are poorly
known for most bursts.
In particular, $n$ could easily lie anywhere in a fairly wide range,
where the majority are within $1 < n < 30$ cm$^{-3}$ \citep{pan01, pan02}.

Following the assumption made by \citet{ghi04b} for most of their
samples, we initially assume $n=3$~cm$^{-3}$ and $\eta_\gamma=0.2$.
Allowing $A$ to vary from 1950~keV to 4380~keV in Eq.~(\ref{eq:e3})
then gives the time interval in which the jet break is expected to
occur if the Ghirlanda relation is satisfied, assuming these values
of $n$ and $\eta_\gamma$ (or equivalently, that $n\eta_\gamma = 0.6$). 
Allowing $n$ to vary between $1-30~{\rm cm}^{-3}$
(or equivalently, $0.2 < n\eta_\gamma < 6$), in Eq.~(\ref{eq:e3})
gives the time interval in which
the jet break is expected to occur if the Ghirlanda relation is
satisfied without assuming a particular value of $n\eta_\gamma$.
The intervals thus obtained are also plotted in
Fig.~\ref{fig:xlc1}--\ref{fig:xlc3}. The dash-dotted, dashed, and
solid lines show the allowed time intervals, without assuming a
particular value of $n\eta_\gamma$ and taking into account the errors in
\Eiso~and \Epsrc; assuming a particular value of $n\eta_\gamma$ and taking
into account the errors in \Eiso~and \Epsrc; and assuming a particular
value of $n\eta_\gamma$ without taking into account the errors in \Eiso~and
\Epsrc.
The time interval in which the jet break is expected to occur was 
completely observed for XRF~050416a and GRB~050525a, but no temporal
break is seen within the interval.
The break at about 11000 sec for GRB~050525a, which is close to the
edge of the expected time interval, was suggested to be
a possible jet break because of its achromatic feature between
X-ray and optical bands \citep{blu06}.
However, if we consider the discrepancy in the spectral and temporal relations
with the theoretical predictions as well, it is suggested that the break is not a jet break.
For GRB~050401, time intervals on
both sides of the time interval were observed and can be joined with a
single power-law decay.
Thus, none of the three bursts exhibit a jet
break within the time period required if they are to satisfy the
Ghirlanda relation.

We now consider the implications if the jet break occurs at either an
earlier or a later epoch than the expected time interval.  If we assume
that the break at $t_{\rm b}$ corresponds to the jet break time, the
temporal decay indices are inconsistent with the values predicted by
the standard afterglow model, as already discussed.  In addition, the
values of $\theta_{\rm j}$ are smaller than their values for other 
bursts.
If we assume, on the other hand, that the jet break occurs after the
time interval covered by the Swift XRT observations, we can derive a
lower limit on $\theta_{\rm j}$, and hence on $E_\gamma$, from the last
time at which the afterglow was detected.  Fig.~\ref{fig:ghi} shows
that in this case the three bursts are also outliers of the
Ghirlanda relation.
Reconciling the X-ray afterglow light curve observed for these three 
bursts with the standard afterglow model requires very unusual values 
of $n$ and/or $\eta_{\gamma}$.  In order to derive from Eq.~\ref{eq:e3}
a jet break time that corresponds to  $t_{\rm b}$, the product of $n$
and $\eta_\gamma$ must be around $200$, $50$ and $10$ for GRB
050401, XRF 050416a, and GRB 050525a, respectively.  In order to derive
from Eq.~\ref{eq:e3} a jet break time that is later than the last
detections, the product of $n$ and $\eta_\gamma$ should be smaller than
$0.2$, $2\times10^{-4}$ and $2\times10^{-3}$ for the three bursts,
respectively \citep[see also][]{lev05}.

For 18 GRBs detected in the pre-\Swift~era, \citet{lia05} found a tight
correlation among \Epsrc, \Eiso, and the rest-frame jet break time 
$t_{\rm jet}^{\rm src}$ \citep[see also][]{xud05}.  The Liang-Zhang
relation is model-independent, while the Ghirlanda relation is not
because the jet opening angle is estimated using the standard jet model
of the afterglow.
If the optical breaks discussed in \citet{lia05} are jet breaks,
the Liang-Zhang relation is equivalent to the
Ghirlanda relation \citep{lia05, xud05, nav06}.
We have shown here
that if we apply a theory of achromatic jet break in the afterglow that was used
in the pre-\Swift~era, the three \Swift~GRBs we have analyzed are
outliers of the Ghirlanda relation.
Therefore, these three \Swift~bursts are also outliers of the Liang-Zhang relation
if the optical breaks discussed in \citet{lia05} are jet breaks.
In fact,
the rest-frame jet break time $t_{\rm jet}^{\rm src}$ can be predicted
using the Eq.~(5) of \citet{lia05} and the values of \Epsrc~and \Eiso~that
we have derived from the observations.  The derived values of $t_{\rm
jet}^{\rm src}$ for the three events are $1-2$ days after the bursts in
the observer frame.  However, no break is visible at that epoch in the
light curves of X-ray afterglows.

Up to now, we have assumed that the ambient density is uniform. 
However, \citet{nav06} have investigated the case of a wind profile
(i.e., $n\propto r^{-2}$) and find that \Egamma~is again tightly
correlated with \Epsrc.  Since this Ghirlanda-wind correlation is also
equivalent to the Liang-Zhang relation \citep{nav06}, the three
\Swift~GRBs discussed in this paper are also outliers of the
Ghirlanda-wind relation.

\subsection{Implications of no Jet Break Feature in the X-ray Band}
 
As discussed in the previous subsection (\S\ref{sec:dis1}), we find
that, for the three bursts we consider, the empirically derived 
Ghirlanda relation is incompatible with the standard jet model of GRB
afterglows that worked well prior to \Swift.  This may be because 
prior to \Swift~jet breaks were observed mainly in the optical band,
whereas in this paper, we have investigated the presence or absence
of jet breaks in the X-ray band.  We consider two possible ways of
reconciling this discrepancy.

One possibility is that the jet break takes place in the optical band
at the time expected from the Ghirlanda relation, even for the three
bursts that we have studied, but that it is masked in the X-ray band by
one or more sources of additional emission, such as (1) inverse Compton
emission, (2) emission from a cocoon around the jet, (3) emission from
the external shock as it passes through a dense region in the
surrounding medium, (4) continuous injection of energy into the
external shock producing a separate source of X-ray emission, or (5) a
separate jet component \citep{pan06}.  If one or more additional
components contribute to the X-ray afterglow emission, one might expect
the observed afterglow to exhibit bumps and/or dips. However, the
observed lightcurves of the X-ray afterglows for the three bursts all
exhibit a rather simple power-law decay.  This, plus the fact that the
decay slopes of the X-ray afterglows of the three bursts that we
consider are shallower both before the observed break and after than in
the cases of many of the light curves of optical afterglows observed
prior to \Swift, favors the possibility that energy is being
continuously injected into the external shock and the X-ray emission
resulting from this injection of energy masks the jet break in X-ray
band that is associated with the jet break in the optical band.

A second possibility is that the jet break occurs at a later time
compared to the time it occurs in previous samples of GRBs.  If this is
case, many \Swift~GRBs would have to belong to a different class of
events from those detected by previous missions, such as {\it CGRO} BATSE,
{\it RXTE}, {\it BeppoSAX} and {\it HETE-2}, and it would imply that the Ghirlanda
relation has a larger scatter than previously thought.  It is difficult
to assess observational selection effects, given the limited number of
\Swift~GRBs for which \Eiso~and \Epsrc~are known as of July 2005.  However,
the properties of the prompt emission of the three bursts that we
consider here are indistinguishable from those of bursts detected prior
to \Swift, and their values of \Epsrc~and \Eiso~satisfy the Amati
relation, making the possibility that many \Swift~GRBs belong to a 
different class of events from those detected by previous missions
seem unlikely. 

It is also possible that both scenarios occur in different bursts.
Simultaneous X-ray and optical observations of GRB afterglows around
the expected jet break time could distinguish between these two 
possibilities.  In the former scenarios, a jet break at the expected
time should be seen in the optical band but not in the X-ray band,
while in the latter scenario, a jet break should not be seen in either
the optical band or the X-ray band.

Current observational evidence is limited and ambiguous.  In the case
of GRB~050401, \citet{pan06} report that the optical lightcurve extends
to $\sim10$ days without a break.  This period covers the entire time
interval during which a jet break is expected, if the burst satisfies
the Ghirlanda relation.  Therefore, either version (5) of the former
scenario, in which the X-ray and the optical afterglows are due to
separate jet components, or the latter scenario are preferable for this
event.  A recent Swift burst, GRB 060206, showed a late-time steepening
of the optical light curve, but no break in the X-ray light curve at
the same time \citep{mon06}, which also lends support to version (5) of
the former scenario, in which the X-ray and the optical afterglows are
due to separate jet components.

In the case of XRF~050416a, there are no observations of the afterglow in
the optical band in the expected time interval.  However, the decay
slope of the afterglow lightcurve of XRF~050416a in X rays is shallow
out to 75 days after the burst \citep{lev06,hul06}.  The X-ray
afterglows of XRF 020427 \citep{ama04} and XRF 050215b \citep{sak06b}
are also shallow out to very late times.  The  properties of the prompt
emission and the shallow decay slopes of the X-ray afterglows imply
that, in the cases of these two bursts, the fluence in the afterglow is
comparable to, or may even exceed, that in the prompt emission
\citep[see also][]{obr06}.
This suggests that the efficiency of the prompt emission may be relatively
small, and provides support for the possibility that energy is being
continuously injected into the external shock, and delays the jet break
in X rays beyond the time of the last XRT observations \citep{sak06b}. 
It also is conceivable that continuous energy injection into the
external shock could power X-ray emission that masks the usual jet
break in X-rays.

In the case of GRB~050525a, \Swift~UVOT in six different filters (V, B,
U, UVW1, UVM2, UVW2) from $\sim$T+70 to $\sim$T+50000~s \citep{blu06}. 
However, only upper limits are available in all the bands after
$\sim$T+50000~s, which unfortunately lies in the middle of the time
interval in which the jet break is expected to occur.  After the source
had faded below the detection limit of UVOT, there are detections in
unfiltered light after $\sim10^5$ s.  However, those observations are not
sufficient to determine if there is a break or not.

The lack of an observed jet break in the X-ray afterglows of the
three bursts we consider here also has implications for the use
of GRBs for cosmological studies.  It has been suggested that the
tightness of the Ghirlanda relation, as reported prior to our study,
makes it possible to use GRBs as ``standard candles'' for constraining
the properties of dark energy \citep{ghi04a}.  Our results suggest 
additional caution in using the Ghirlanda relation for this purpose.

Given the importance of the presence or absence of a jet break for
understanding the nature of GRB jets and for the use of GRBs as
``standard candles'' for cosmology, we strongly encourage simultaneous
X-ray and optical observations of GRB afterglow around the expected jet
break time for GRBs having reliable measurements of \Epobs~and
redshifts in order to investigate whether the breaks seen in the
optical are accompanied by breaks in the X-ray at the time expected in
the standard jet model of GRB  afterglows.

\acknowledgments
We gratefully acknowledge support from the \Swift~team and the
Konus-Wind team.
G. S. is supported by
the Research Fellowships for Young Scientists (2002--2005)
and the Postdoctoral Fellowships for Research Abroad (2006--)
of the Japan Society for the Promotion of Science.
T. S. is supported by the NASA Postdoctoral Program
administered by Oak Ridge Associated Universities at NASA Goddard Space
Flight Center.  This research is supported in part by Grants-in-Aid for
Scientific Research of the Japanese Ministry of Education, Culture,
Sports, Science, and Technology 18740153(R.~Y.), 14079207(T.~T.), 14047212 (T.~N.),
14204024 (T.~N.).

\clearpage

\begin{deluxetable}{cccccc}
\tabletypesize{\scriptsize}
\rotate
\tablecaption{Redshifts, durations, and emission properties of three \Swift~GRBs\label{tab:sample}}
\tablewidth{0pt}
\tablehead{
\colhead{GRB} &
\colhead{redshift} &
\colhead{$T_{90}$} &
\colhead{Peak 1 s Flux in 15-150 keV} &
\colhead{Fluence in 15-150 keV} &
\colhead{Observing Instruments}\\
 &
 &
\colhead{[s]} &
\colhead{[erg cm$^{-2}$ s$^{-1}$]} &
\colhead{[erg cm$^{-2}$]} & 
}
\startdata
\\
050401 &
2.9 &
34.3 &
$9.26_{-0.72}^{+0.72}\times 10^{-7}$ &
$8.49_{-0.32}^{+0.32}\times 10^{-6}$ &
\Swift~BAT, \Swift~XRT, VLT, \\
 & & & & &
 Optical, ROTSE, Siding Spring\\
 \\
050416a &
0.65 &
2.4 &
$2.02_{-0.20}^{+0.20}\times 10^{-7}$ &
$3.36_{-0.32}^{+0.34}\times 10^{-7}$ &
\Swift~BAT, \Swift~XRT, Keck\\
\\
050525a &
0.606 &
8.9 &
$3.62_{-0.06}^{+0.06}\times 10^{-6}$ &
$1.62_{-0.02}^{+0.2}\times 10^{-5}$ &
\Swift~BAT, \Swift~XRT, \Swift~UVOT, VLA\\
& & & & &
INTEGRAL, Spitzer, HST, Numerous Optical \& NIR\\
\\
\enddata
\end{deluxetable}

\clearpage

\begin{deluxetable}{ccccccccccccc}
\tabletypesize{\scriptsize}
\rotate
\tablecaption{Prompt emission spectral parameters and 
X-ray afterglow temporal and spectral parameters
\label{tab:parms}}
\tablewidth{0pt}
\tablehead{
\colhead{GRB} &
\colhead{$\alpha_{\rm B}$} &
\colhead{$\beta_{\rm B}$} &
\colhead{\Epobs~[keV]} &
\colhead{\Epsrc~[keV]} &
\colhead{\Eiso~[$10^{52}$ erg]} &
\colhead{$t_{\rm b}$ [s]} &
\colhead{$\alpha_{1}$} &
\colhead{$\alpha_{2}$} &
\colhead{$\Gamma_{1}$} &
\colhead{$\Gamma_{2}$}
}
\startdata
\\
050401 &
$-0.87_{-0.27}^{+0.36}$ &
$-2.47_{-0.36}^{+0.21}$ &
$115_{-16}^{+19}$ &
$447_{-64}^{+75}$ &
$34.3_{-3.4}^{+3.7}$ &
$5390\pm450$ &
$-0.57\pm0.02$ &
$-1.34\pm0.05$ &
$2.03_{-0.05}^{+0.05}$ &
$1.98_{-0.24}^{+0.26}$ \\
\\
050416a &
- &
$<-3.35$ &
$17.3_{-8.0}^{+3.0}$ &
$28.5_{-13.2}^{+5.0}$ &
$0.083_{-0.013}^{+0.065}$ &
$1670\pm600$ &
$-0.55\pm0.06$ &
$-0.82\pm0.02$ &
$2.20_{-0.24}^{+0.27}$ &
$2.04_{-0.15}^{+0.16}$ \\
\\
050525a &
$-0.97_{-0.10}^{+0.11}$ &
$<-2.55$ &
$78.2_{-1.6}^{+4.7}$ &
$125.6 _{-2.6}^{+7.6}$ &
$2.23_{-0.11}^{+0.03}$ &
$10600\pm3300$ &
$-1.18\pm0.02$ &
$-1.51\pm0.06$ &
$1.92_{-0.05}^{+0.05}$ &
$2.11_{-0.39}^{+0.28}$ \\
\\
\enddata
\tablecomments{$\alpha_{\rm B}$, $\beta_{\rm B}$, \Epobs~and \Epsrc are the parameters of the Band function.  \Eiso~is calculated in $1$--$10^4$ keV.  $t_{\rm b}$ is break time observed in X-ray band. $\alpha_{1}$ and $\alpha_{2}$ are decay indices of the X-ray afterglows before and after $t_{\rm b}$.
$\Gamma_{1}$ and $\Gamma_{2}$ are photon indices of the X-ray afterglows before and after $t_{\rm b}$.}
\end{deluxetable}

\clearpage

\clearpage

\clearpage

\clearpage

\clearpage

\clearpage

\end{document}